\documentclass{epl}
\usepackage{graphicx}
\usepackage{amssymb}
\usepackage{amsmath}

\def\ba{\begin{eqnarray}}
\def\ea{\end{eqnarray}}
\def\beq{\begin{eqnarray}}
\def\eeq{\end{eqnarray}}
\def\be{\begin{equation}}
\def\ee{\end{equation}}

\title{Domain Growth in Random Magnets}
\author{Raja Paul\inst{1}, Sanjay Puri\inst{2} and Heiko Rieger\inst{1}}
\institute{
\inst{1}Theoretische Physik, Universit\"at des Saarlandes-66041 Saarbr\"ucken, GERMANY.\\
\inst{2}School of Physical Sciences, Jawaharlal Nehru University, New
Delhi-110067, INDIA.}


\begin{document}

\maketitle

\begin{abstract}
We study the kinetics of domain growth in ferromagnets
with random exchange interactions. We present detailed
Monte Carlo results for the nonconserved
random-bond Ising model, which are consistent with power-law
growth with a variable exponent. These results are interpreted in
the context of disorder barriers with a logarithmic dependence on
the domain size. Further, we clarify the implications of logarithmic
barriers for both nonconserved and conserved domain growth.
\end{abstract}

A homogeneous binary mixture becomes thermodynamically unstable if it
is rapidly quenched below the coexistence curve.
The subsequent far-from-equilibrium
evolution of the system is characterized by the emergence and growth
of domains enriched in the new equilibrium phases. The domain
morphology is quantified by (a) the time-dependence of the domain
scale $R(t)$, where $t$ is the time after the quench; and (b)
the correlation function or its Fourier transform, the structure
factor \cite{ab94}. There is a good understanding of domain-growth
kinetics in pure and isotropic systems, where the domain scale
shows a power-law behavior, $R(t) \sim t^\theta$.
For the case with nonconserved order parameter, e.g., ordering
of a ferromagnet into up and down phases, we have $\theta = 1/2$. On
the other hand, for the case with conserved order parameter, e.g., phase
separation of a binary (AB) mixture into A- and B-rich domains, we
have $\theta = 1/3$ when growth is driven by diffusion.

Of course, real experimental systems are neither pure nor isotropic. In this
letter, we focus on domain growth in ferromagnets and binary alloys with
quenched disorder in the form of random exchange interactions. There
have been many experimental \cite{iei90,saw93,lla00} and numerical
\cite{gs85,oc86,cgg87,pcp91,bh91,ghss95} studies of this problem
\cite{sp04}. At early times, domain coarsening is not affected by
disorder. Then, there is a crossover to a disorder-affected regime,
which occurs earlier for higher disorder amplitudes.
There have also been many studies of domain growth in {\it spin glasses}
\cite{bckm97,naoki_rieger}, where the amplitude of disorder is such that
the local exchange coupling may be either ferromagnetic or antiferromagnetic.
Inspite of this attention, the nature of asymptotic domain growth in both
random magnets and spin glasses remains the subject of much controversy.
The present letter resolves this controversy in the context of random magnets.
We present detailed Monte Carlo (MC) results which show that asymptotic
growth in these systems is consistent with a power-law behavior,
with the growth exponent $(\theta)$ depending on the
temperature ($T$) and disorder amplitude ($\epsilon$). Further, we present
analytical arguments which clarify the functional form of $\theta (T,\epsilon)$
for systems with both nonconserved and conserved order parameters.

We study the random-bond Ising model (RBIM) with the following Hamiltonian:
\begin{equation}
\label{rbim}
H = - \sum\limits_{\langle ij \rangle} J_{ij} S_i S_j, \quad S_i = \pm 1,
\end{equation}
where the exchange couplings $J_{ij}$ are drawn from a probability
distribution, e.g., Gaussian, uniform, bimodal, etc. We focus on
the nearest-neighbor case, denoted by the subscript $\langle ij \rangle$. 
A kinetic version of the RBIM is obtained by associating Glauber spin-flip
kinetics $(S_i \rightarrow -S_i)$ or Kawasaki spin-exchange
kinetics $(S_i \leftrightarrow S_j)$ with the Hamiltonian in Eq.~(\ref{rbim}).
In this letter, we consider the ferromagnetic case, where $J_{ij} > 0$
always. The case where $J_{ij}$ can be both $> 0$ (ferromagnetic) and $< 0$
(antiferromagnetic) is relevant to spin glasses.

An important study of the nonconserved RBIM is due to Huse and
Henley (HH) \cite{hh85}. HH argued that coarsening
domains are trapped by energy barriers $E_B(R) \simeq E_0 R^{\psi}$,
with exponent $\psi = \chi/(2- \zeta)$, where
$\chi$ and $\zeta$ are the pinning and roughening exponents. For $d=2$, these
exponents are known to be $\chi = 1/3$ and $\zeta=2/3$ \cite{fns77},
yielding $\psi=1/4$. For $d=3$, a perturbative calculation gives $\psi \simeq 0.55$
\cite{hh85}. To obtain the corresponding growth law, it is convenient to
use the framework of Lai et al. (LMV) \cite{lmv88}. For curvature-driven growth
in nonconserved systems, the length scale obeys
\begin{equation}
\label{curv}
\dot{R} = a(R,T)/R ,
\end{equation}
where $a(R,T)$ is the diffusion constant which depends (in general) upon
both the length scale and the temperature. In the presence of energy barriers,
domain growth proceeds via thermally-activated barrier hopping with $a(R,T) \simeq
a_0 \exp (-E_B/T)$. Replacing the HH energy-barrier scaling in
Eq.~(\ref{curv}), we obtain the crossover behavior:
\begin{equation}
\label{cross}
R(t) \simeq R_0(T,\epsilon) h \left( \frac{t}{t_0} \right) , \quad \rm{with}
\end{equation}
\beq
\label{hh}
R_0(T,\epsilon) &=& \left( \frac{T}{E_0} \right)^{1/\psi} ,
\quad t_0(T,\epsilon) = \frac{1}{a_0 \psi} \left( \frac{T}{E_0}
\right)^{2/\psi}, \quad \rm{and}
\eeq
\beq
h(x) &=& \left( \frac{2}{\psi} x \right)^{1/2}, \quad x \ll 1, \nonumber \\
&=& \left( \ln x \right)^{1/\psi}, \quad x \gg 1 .
\eeq

The logarithmic growth law proposed by HH has motivated many experimental
and numerical studies of domain growth in random magnets. However, to
date, there is no clear confirmation of this growth law. For example,
the experiments of Ikeda et al. \cite{iei90} found that there is no
universal logarithmic law. Numerical studies have also found no evidence for
a HH-type law, valid over extended time-windows and parameter
regimes. On the contrary, some experiments \cite{lla00} and simulations
\cite{oc86} suggest power-law growth with a variable exponent.

Let us next present results from our MC simulations, which were done on
$L \times L$ lattices (in $d=2$) with periodic boundary conditions. The results
presented here were obtained for uniform $J_{ij}$-distributions on
the interval $[1-\epsilon/2, 1+\epsilon/2]$ or $[1-\epsilon', 1]$, where
$\epsilon$ and $\epsilon'$ quantify the degree of disorder. The limits
$\epsilon = 0$ (or $\epsilon' = 0$) and $\epsilon = 2$ (or $\epsilon' = 1$)
correspond to zero and maximum disorder, respectively. Assigning
random initial orientations to each spin, we rapidly quench to
$T<T_c \simeq 2.269$. All our statistical data is obtained with lattice sizes
$L=1024$, as an average over 50 independent initial conditions and
disorder configurations. We also produced data for $L = 256$ and
$512$ (not shown) to ensure that our results are not influenced by
finite size effects.

\begin{figure}
\onefigure[scale=0.55]{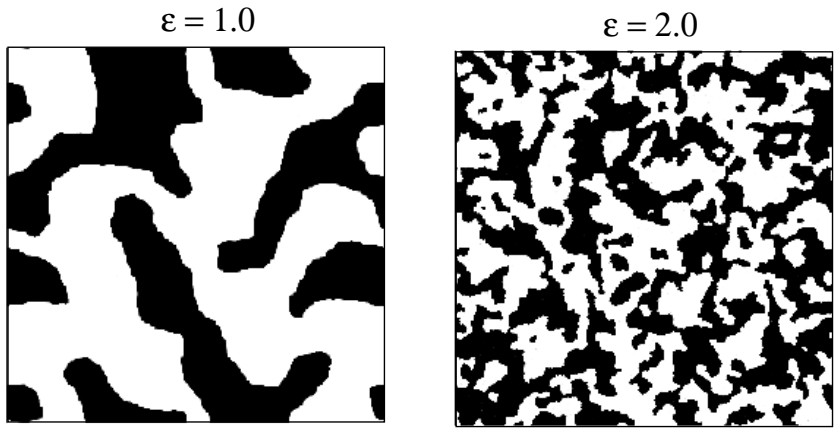}
\caption{Domain growth in the RBIM with Glauber kinetics. We show evolution
pictures at $t=10^{5}$ MCS for a $512 \times 512$ lattice, after a quench
from $T = \infty$ to $T = 0.5$. The up spins are marked in black, and the
down spins are unmarked. The snapshots correspond to different
disorder amplitudes, $\epsilon = 1,2$.}
\label{fig1}
\end{figure}

Figure~\ref{fig1} shows typical evolution snapshots (at $t = 10^5$ MCS) for
different disorder amplitudes $\epsilon = 1,2$. As expected, the
length scale at a given time decreases with increase in $\epsilon$.
However, the domain morphology does not visually differ from that for
the pure case. We have confirmed (not shown here) that the scaled correlation
function [$C(r,t)$ vs. $r/R(t)$] is independent of the disorder
amplitude, and is numerically equivalent to that for the pure
case $(\epsilon = 0)$. This observation has earlier been made by
Puri et al. \cite{pcp91} and Bray and Humayun \cite{bh91}.

\begin{figure}
\twoimages[scale=0.28]{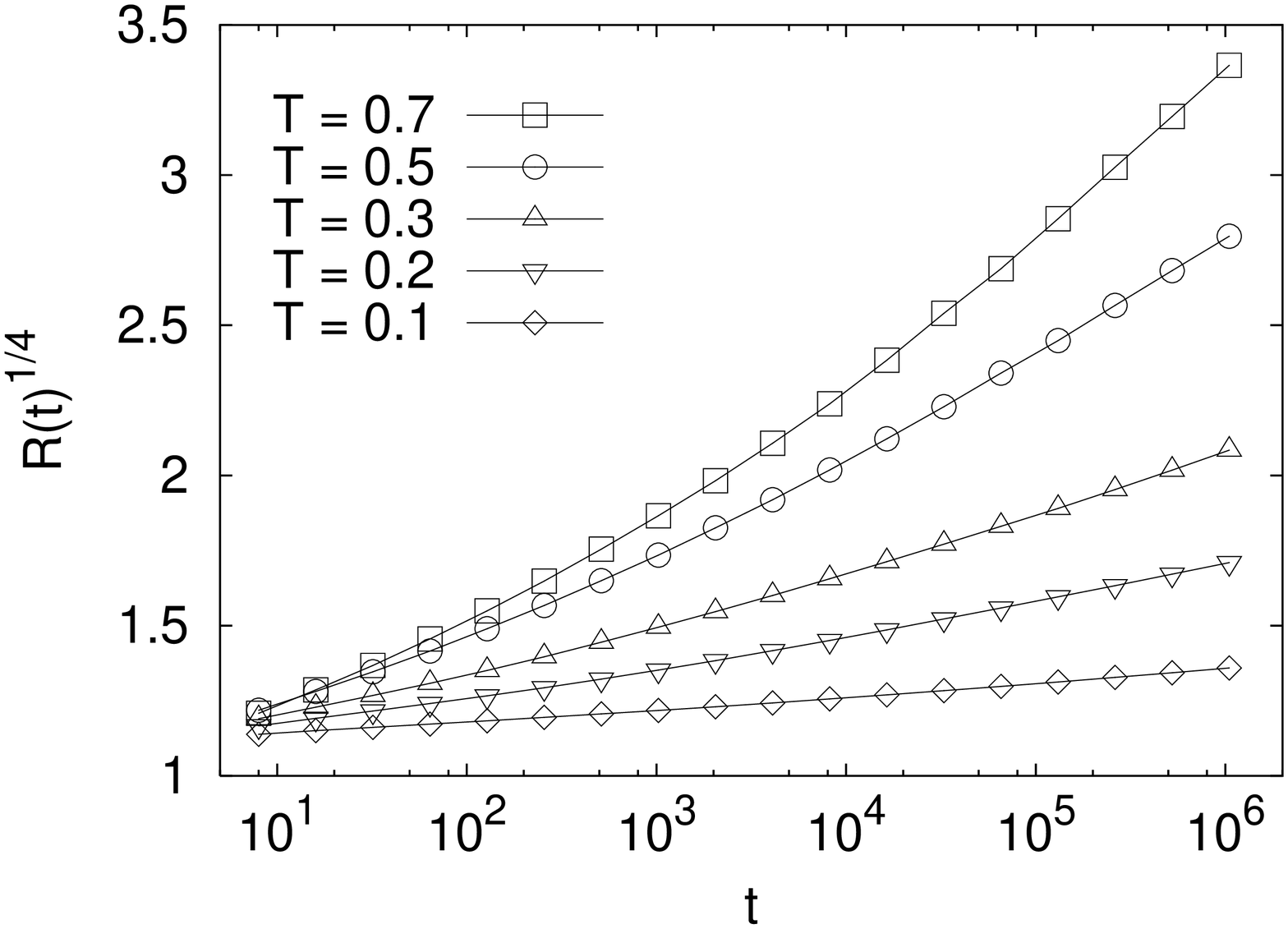}{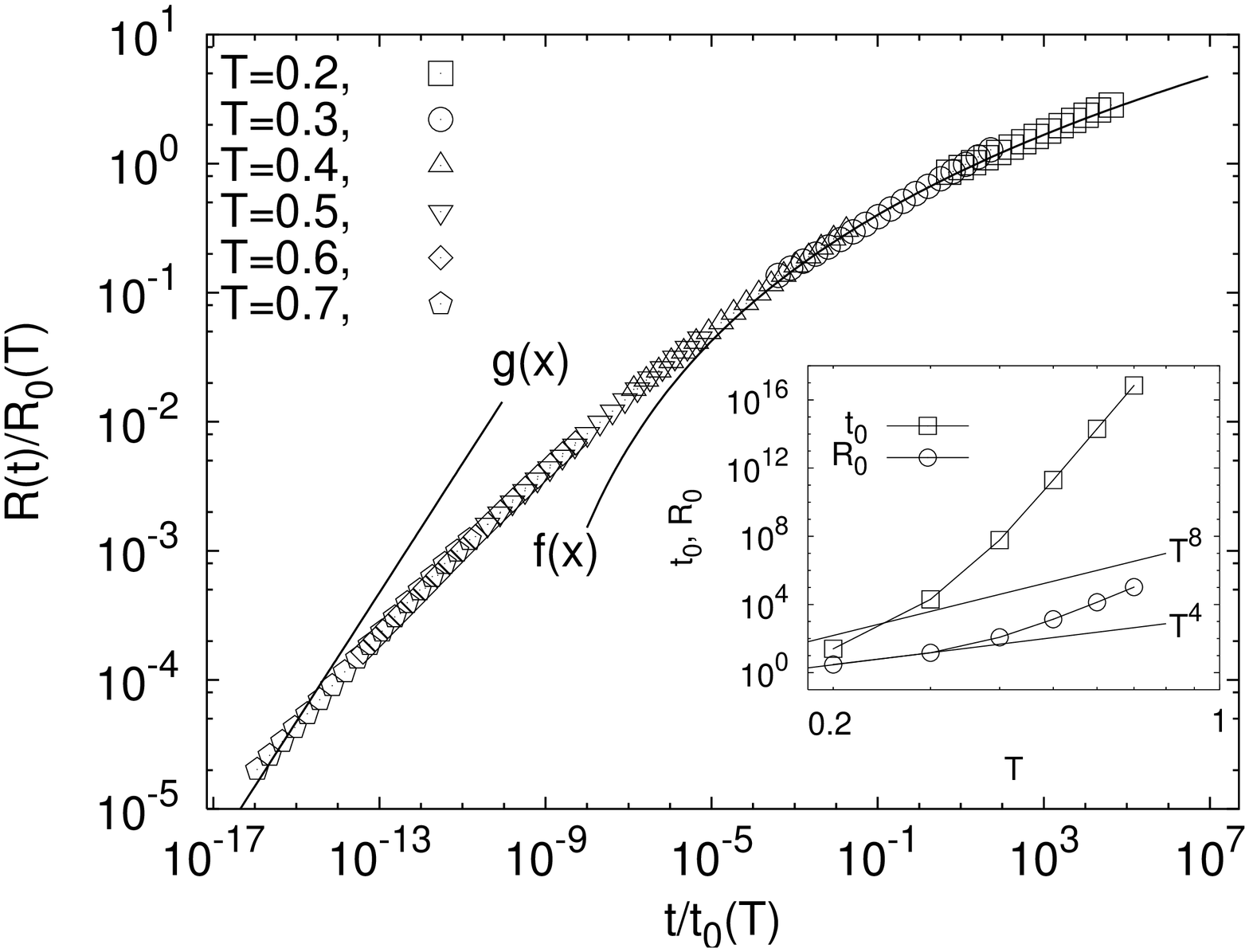}
\caption{{\bf Left:} Plot of $R^{1/4}$ vs. $t$ on a log-linear scale for
$\epsilon' =1$, i.e. $J_{ij}$ is uniformly distributed between
$[0,1]$, and different quench temperatures. We show data for 5
temperature values: $T = 0.1,0.2,0.3,0.5,0.7$. {\bf Right:} Scaling
plot according to Eqs.~(3)-(5) ($J_{ij}$ as in left figure). For each
temperature $T$ the values for $R_0(T)$ and $t_0(T)$ have been chosen
so as to obtain a smooth scaling curve $h(x)$. The
functions $g(x)\propto x^{1/2}$ and $f(x)\propto(\ln x)^4$ represent
the expected asymptotic behaviors for $x \ll 1$ and $x \gg 1$,
respectively, according to Eq.~(5). The inset shows the temperature-dependence
of the fit values $R_0(T)$ and $t_0(T)$ and their expected
$T$-dependence according to Eq.~(4), which is $T^4$ and $T^8$,
respectively, if $a_0$ and $E_0$ are only weakly (sublinearly) 
$T$-dependent.}
\label{fig2}
\end{figure}

Next, we consider the time-dependence of the domain size, which is
obtained as the distance over which the correlation function decays to
half its maximum value. In Fig.~\ref{fig2}, we undertake a direct test
of the HH law by first plotting. $R^{1/4}$ vs.  $t$ on a log-linear
scale for different quench temperatures and $\epsilon' = 1$, with
$J_{ij} \in [1-\epsilon',1]$. The data does not satisfy the asymptotic
HH growth law, which corresponds to a straight line on this plot.
More generally, we have attempted to fit the data to the logarithmic
function, $\ln t=aR^x+b$. This function does not give a good fit to
the data.  For acceptable fits, the exponent $x$ depends strongly on
the temperature (cf. Ref.~\cite{iei90}), in contrast with the
prediction of a universal growth law. Similar results are obtained if
we fix $T$ and vary $\epsilon$, and we do not present these here. 

We also tried to fit the data sets to the crossover scaling form
described by Eqs.~(3)-(5), the result for which is shown in the right
portion of Fig.~2. We record the following points of disagreement with
the proposed scaling: (1) The short-time behavior is not
described well by Eq.~(5), i.e., in the plot $g(x)$ does not fit
well to the scaling curve. (2) The proposed asymptotic behavior,
i.e., the curve $f(x)$, does not fit the scaling curve well even
for the largest times. (3) The temperature-dependence of the crossover
length $R_0(T)$ and the crossover time $t_0(T)$ is stronger than a
power law, which is incompatible with the expected behavior in Eq.~(4).
The parameters $E_0$ and $a_0$ occurring in Eq.~(4) are
expected to decrease with increasing temperature, and therefore $R_0$
and $t_0$ may be expected to increase faster than $T^4$ and $T^8$,
respectively. However, their putative $T$-dependence turns out to be much too strong.
Note that the crossover time $t_0$ in the inset of Fig.~2 varies over 20
decades when $T$ varies over only half a decade from 0.2 to 0.7, and
we do not see why the pinning energy amplitude $E_0$ or the
domain wall tension should have such a strong $T$-dependence.
Based on observations 1-3, we believe that the crossover scaling form
in Eqs.~(3)-(5) does not describe the data well and we suggest here
an alternative picture.

\begin{figure}
\twoimages[scale=0.28]{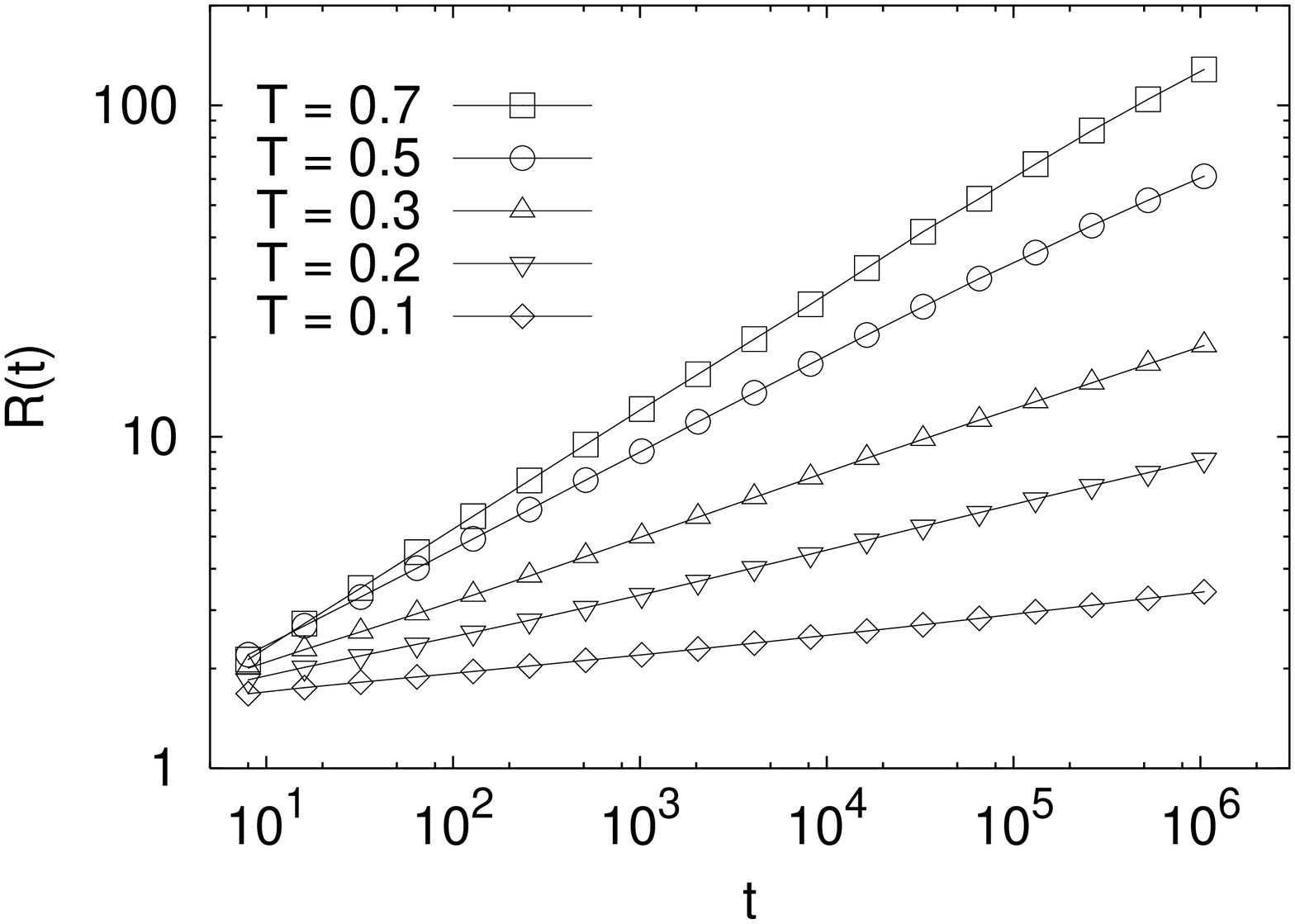}{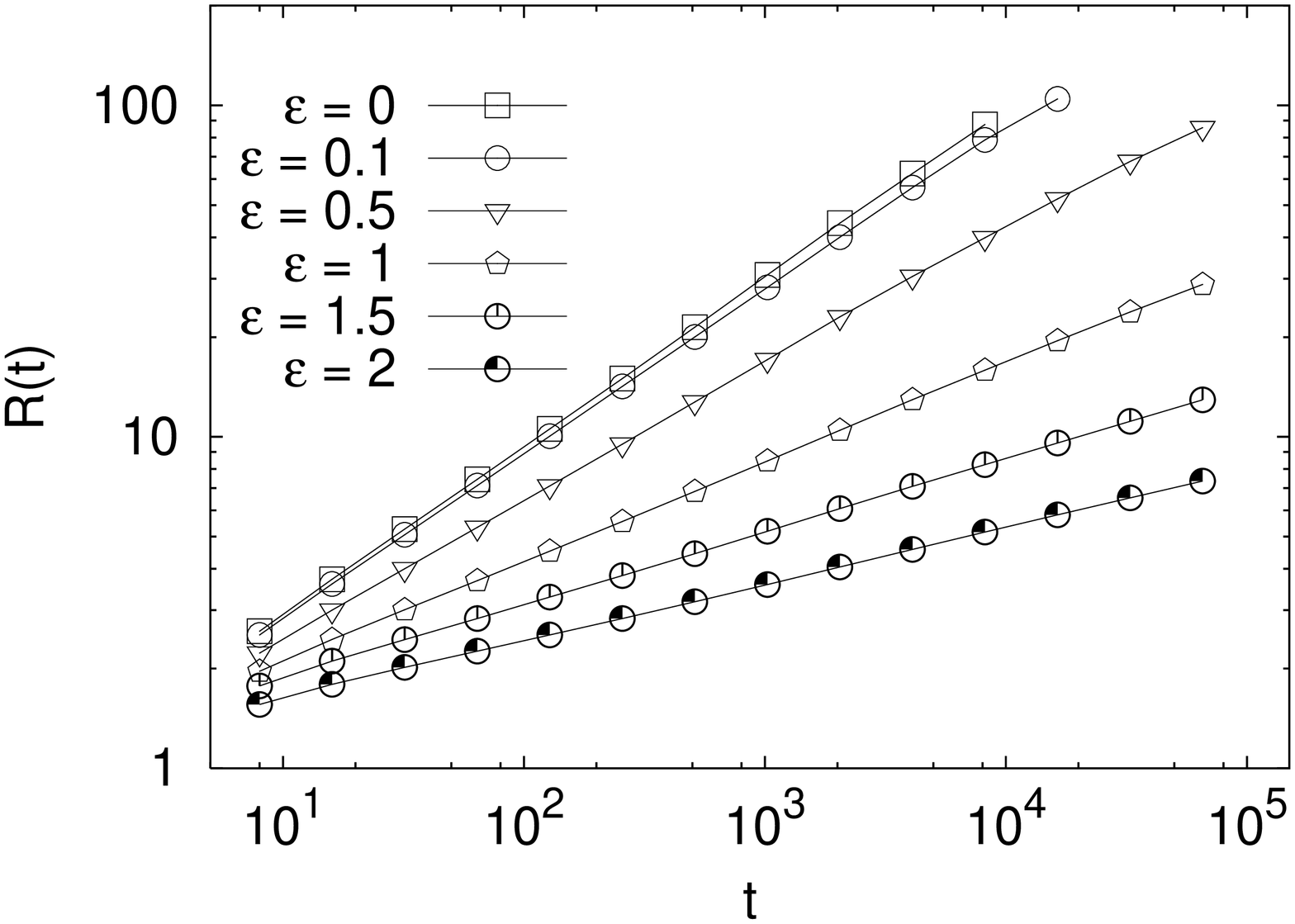}
\caption{{\bf Left:} 
Log-log plot of $R$ vs. $t$ for the length-scale data shown in Fig.~\ref{fig2}. 
{\bf Right:} Log-log plot of $R$ vs. $t$ for $T=0.5$ and different disorder 
amplitudes: $\epsilon = 0$ (pure case), and $\epsilon = 0.1,0.5,1,1.5,2$. 
Here $J_{ij}$ is uniformly distributed over $[1-\epsilon/2, 1+\epsilon/2]$.}
\label{fig3}
\end{figure}

Several groups \cite{lla00,oc86} have reported that random magnets
exhibit power-law growth with variable exponents. In Fig.~\ref{fig3} (left), we plot
$R$ vs. $t$ on a log-log scale for the data in Fig.~\ref{fig2}. For low $T$, the plots
are linear over 5 decades in time. For high $T$, the log-log plots exhibit a
crossover behavior. In Fig.~\ref{fig3} (right), we present a similar plot for $T=0.5$,
and different values of the disorder. Again, the data exhibits a
power-law behavior for large values of $\epsilon$, and crossover
behavior for small values of $\epsilon$. Our data is consistent with
power-law growth with a variable exponent, at least for low $T$ and
high $\epsilon$. We have obtained similar results for domain growth
in the dilute Ising model (DIM), and will present these elsewhere \cite{ppr04}.

Let us understand the origin of growth exponents which depend on
$T$ and $\epsilon$. For the DIM,
Henley \cite{ch85} and Rammal and Benoit \cite{rb85} have argued that
the fractal nature of domain boundaries results in a logarithmic
$R$-dependence of trapping barriers. We propose that this is generally
applicable\cite{remark} and examine the implications thereof. At early times and small
length scales, we expect disorder-free domain growth.
This suggests the barrier-scaling form:
\ba
\label{blog}
E_B(R) \simeq \epsilon \ln \left( 1 + R \right) ,
\ea
where $R$ is measured in dimensionless units. Substituting $a(R,T) \simeq a_0
\exp (-E_B/T)$ in Eq.~(\ref{curv}), we obtain
\beq
\label{log} 
\dot{R} = a_0 (1+R)^{-\epsilon/T}/R .
\eeq 

The solution of Eq.~(\ref{log}) is the crossover function
\ba 
\label{hrb} 
R(t) &\simeq& (2 a_0 t)^{1/2}, \quad t \ll t_0, \nonumber \\
&\simeq& \left[ (2+\epsilon/T) a_0 t \right]^{\theta (T,\epsilon)},
\quad t \gg t_0,
\ea 
with the growth exponent
\ba
\theta (T,\epsilon) = (2+\epsilon/T)^{-1} .
\ea
The early- and late-time behaviors in Eq.~(\ref{hrb}) arise in the limits
$R \ll 1$ and $R \gg 1$, respectively.

The crossover length and time can be identified by reformulating Eq.~(\ref{hrb})
as Eq.~(\ref{cross}) with
\ba
R_0 (T,\epsilon) = {1 \over (2 \theta)^{\theta/(1-2\theta)}},\qquad
t_0 (T,\epsilon) = {1 \over a_0} \frac{1}{(2 \theta^{2\theta})^{1/(1-2 \theta)}},
\quad \rm{and}
\ea
\ba
h(x) &=& x^{1/2}, \quad x \ll 1, \nonumber \\
&=& x^{\theta}, \quad x \gg 1 .
\ea

\begin{figure}
\twoimages[scale=0.28]{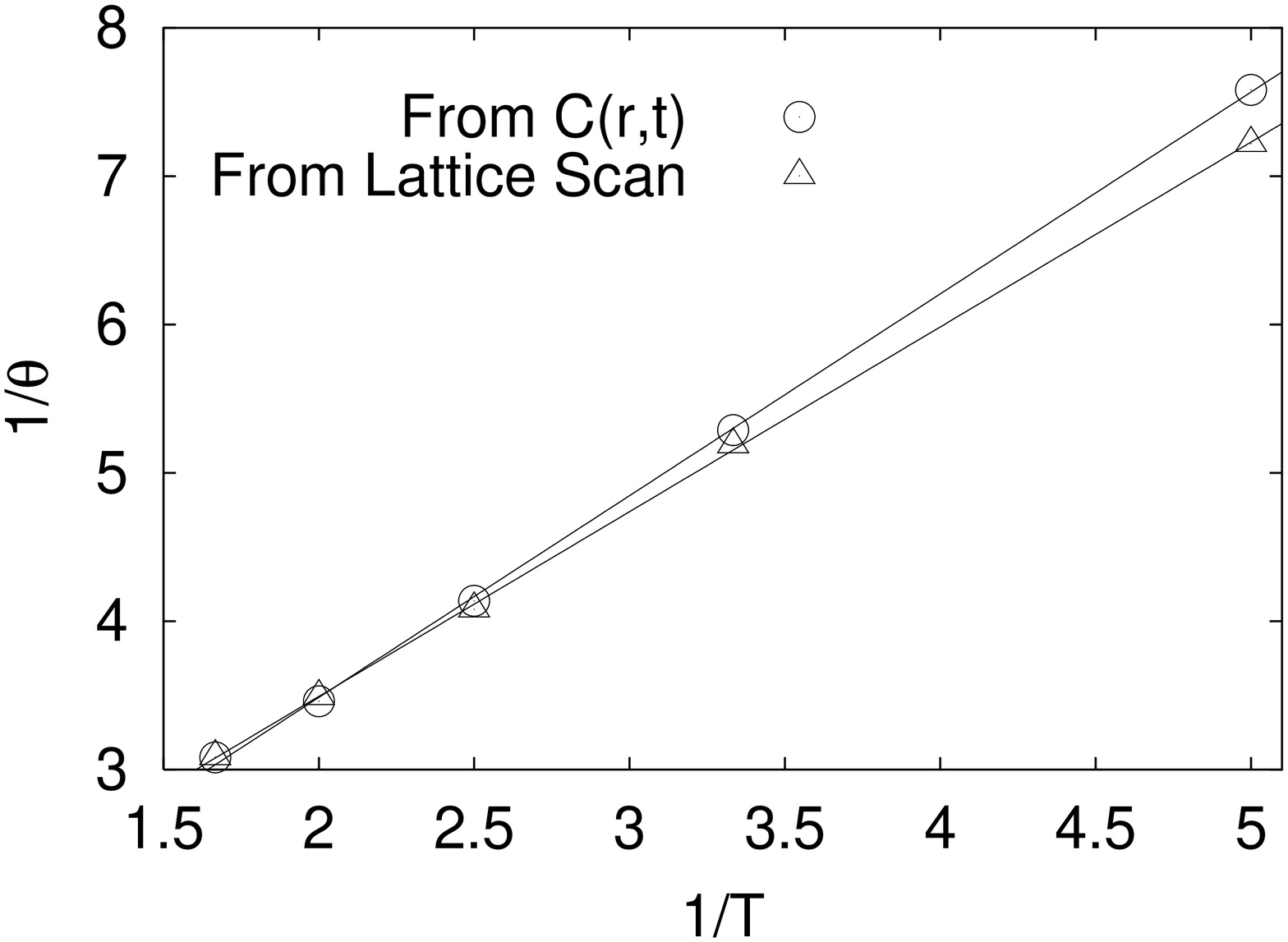}{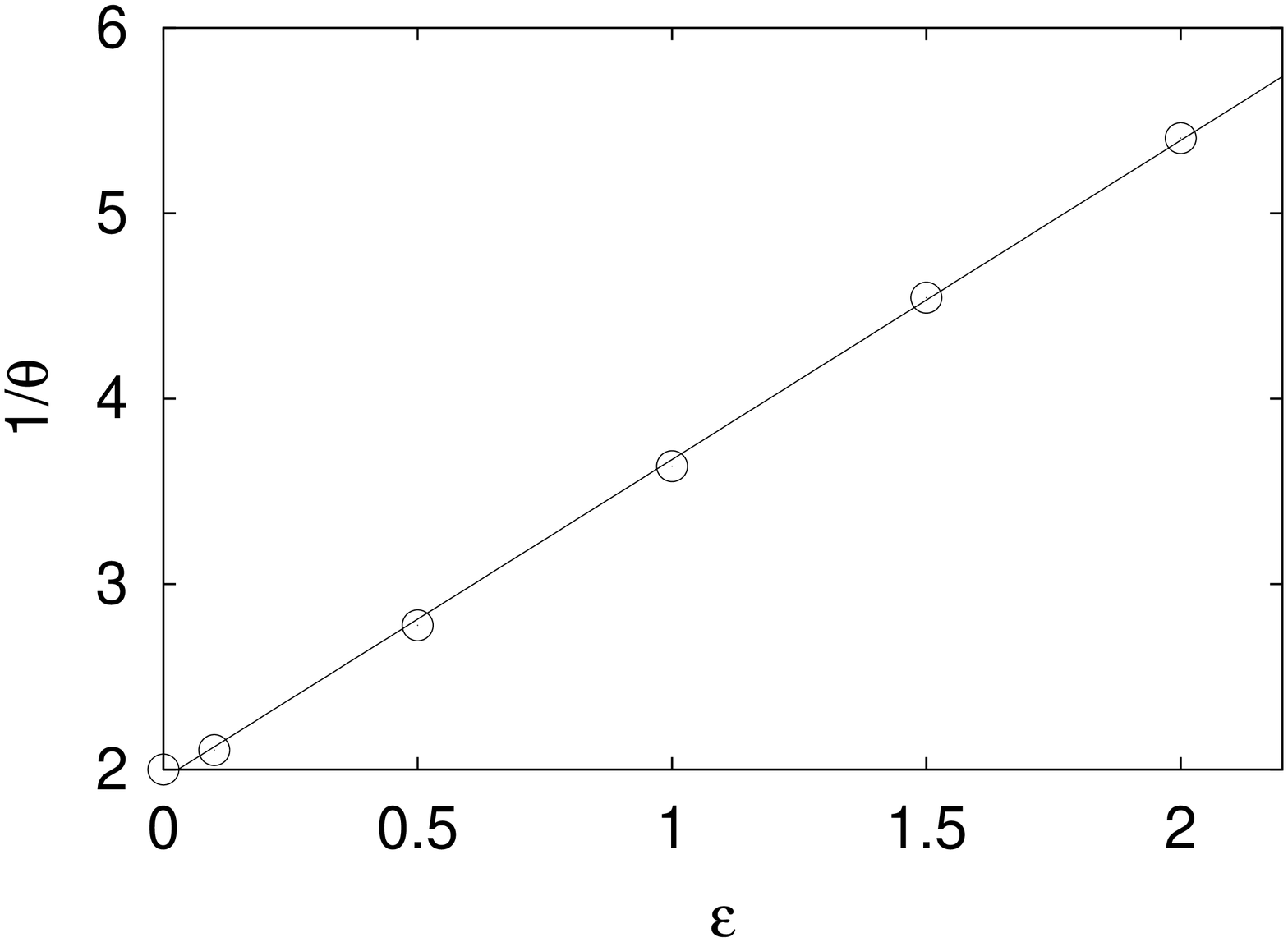}
\caption{{\bf Left:} Exponent $1/\theta$ vs. $1/T$ for the data shown
in the left part of Fig.~\ref{fig3} (circles), and obtained for the
same systems using estimates for the domain size via the lattice
scaling method described in the text (squares). Note that for
$T > 1 \approx T_c(\epsilon'=1)$, i.e., for $1/T<1$ the system is in the
paramagnetic phase. Hence $R$ cannot grow algebraically any more and
no data for $1/\theta$ are available in this region. {\bf Right:}
Exponent $1/\theta$ vs. $\epsilon$ for the data shown in the right
part of Fig.~\ref{fig3}.}
\label{fig4}
\end{figure}

In Fig.~\ref{fig4}, we plot $1/\theta$ vs. $1/T$ and $\epsilon$
for the data shown in Fig.~\ref{fig3}.  The
resultant linear plots confirm the logarithmic barrier-scaling
scenario proposed above. For high temperatures and weak disorder
amplitudes, recall that the length-scale data exhibits a crossover
behavior. In these cases, we estimate the asymptotic exponent by
tracking the effective exponent $\theta_{\rm eff} = d(\ln R)/d (\ln t)$
as a function of time.

We note the following points: (1) When measuring the growth by
alternative methods, we get the same estimates for the growth
exponents. For instance, we applied the following Monte Carlo
renormalization group approach. First, we scaled the lattice twice
by taking 4 block spins to reduce thermal noise. As a result,
a $1024^2$ lattice becomes a $256^2$ lattice.  Then we scanned horizontally and
vertically through the lattice and measured the total number of domain
walls. Finally, we divided the system area by this number to get the average
domain length. From the time-dependence, we extracted the
exponents $\theta$ shown in Fig.~\ref{fig4} (left), and we observe that
they agree well within numerical accuracy. (2) We checked that the
correlation functions that we calculated obey the expected dynamical
scaling behavior: $C(r,t)=c[r/R(t)]=c[r/t^{\theta(T,\epsilon)}]$.

Given our results in the nonconserved case, we expect that the
length-scale data for the conserved case should also exhibit power-law
behavior with a variable exponent. We have undertaken MC simulations
of the RBIM with conserved kinetics, and the results will be presented
elsewhere\cite{ppr04}.  Here, we confine ourselves to discussing the
implications of logarithmic energy barriers for growth exponents in
conserved systems.

In the absence of disorder, the domain scale obeys the Huse equation
\cite{dh86} $\dot{R} = D_0/R^2$, 
with the solution $R(t) \simeq (3D_0 t)^{1/3}$, which is referred to
as the Lifshitz-Slyozov growth law. As before, the presence of disorder
renormalizes the diffusion constant $D_0$ by an Arrhenius factor. For
logarithmic barriers as in Eq.~(\ref{blog}), the corresponding
kinetic equation is
\ba
\label{husel}
\dot{R} = D_0 \left( 1 + R \right)^{-\epsilon/T}/R^2 .
\ea
The short- and long-time solutions of Eq.~(\ref{husel}) are obtained as
follows: $R(t)\simeq(3D_0 t)^{1/3}$ for $t \ll t_0$ and 
$R(t)\simeq\left[ (3+\epsilon/T) D_0 t \right]^{\theta (T,\epsilon)}$
for $t\gg t_0$, where
\ba
\theta (T,\epsilon) = (3+\epsilon/T)^{-1} .
\ea

The asymptotic exponent differs from that for the nonconserved case. This should
be contrasted with the HH scenario, where the asymptotic growth law is the
same for the nonconserved and conserved cases \cite{pcp91,ab94}, which is seen
by incorporating the HH barrier-scaling form in $\dot{R} = D_0/R^2$. 
The crossover form of the solution of Eq.~(\ref{husel}) is Eq.~(\ref{cross}) with 
$R_0 (T,\epsilon) = 1/(3\theta)^{\theta/(1-3\theta)}$ and 
$t_0 (T,\epsilon) = D_0^{-1}/(3\theta^{3\theta})^{1/(1-3\theta)}$
and $h(x)=x^{1/3}$ for $x \ll 1$ and $h(x)=x^{\theta}$ for $x \gg 1$.

In conclusion, we have presented detailed MC results for domain growth
in random magnets for a wide range of temperatures and disorder amplitudes.
Our results do not support the HH scenario of logarithmic domain
growth. (Of course, we cannot rule out the possibility of a
{\it logarithmic} regime at even later times than those investigated
here.) Rather, they are in agreement with results \cite{lla00,oc86}
which exhibit power-law growth with a variable exponent. This scenario
arises naturally in the context of logarithmic energy barriers, and the
corresponding functional dependence of $\theta$ on $T$ and $\epsilon$
is in excellent agreement with our numerical results. The results in this letter
provide a framework for the analysis of experiments and simulations on domain
growth in disordered magnets and binary mixtures.
 
\acknowledgements
This work was financially supported by the Deutsche Forschungsgemeinschaft
(DFG), SFB277.


\end{document}